\documentclass{tMOP2e}

\usepackage{graphicx}
\usepackage{enumerate}
\usepackage{siunitx}
\usepackage{braket}
\usepackage{nicefrac}
\usepackage{xcolor}
\usepackage{upgreek}

\begin{document}
\doi{10.1080/0950034YYxxxxxxxx}
 \issn{1362-3044}
\issnp{0950-0340} \jvol{00} \jnum{00} \jyear{2010} \jmonth{10 January}

    
\title{Sideband cooling of small ion Coulomb crystals in a Penning trap}

\author{G. Stutter$^1$\thanks{$^1$Present Address: Department of Physics and Astronomy, Aarhus University, DK-8000 Aarhus C, Denmark.} , P. Hrmo, V. Jarlaud , M. K. Joshi,  J. F. Goodwin$^2$\thanks{$^2$Present Address: Department of Physics, University of Oxford, Clarendon Laboratory, Parks Road, Oxford OX1 3PU, United Kingdom} and R. C. Thompson$^{\ast}$\thanks{$^\ast$Corresponding author. Email: r.thompson@imperial.ac.uk
\vspace{6pt}}\\\vspace{6pt}  {\em{Quantum Optics and Laser Science, Blackett Laboratory, Imperial College London, Prince Consort Road, London, SW7 2AZ, United Kingdom} }
\\\vspace{6pt}\received{v8 released \today}}
    
\maketitle
    
\begin{abstract}

We have recently demonstrated the laser cooling of a single $^{40}$Ca$^+$ ion to the motional ground state in a Penning trap using the resolved-sideband cooling technique on the electric quadrupole transition  S$_{\nicefrac{1}{2}} \leftrightarrow$ {D}$_{\nicefrac{5}{2}}$. Here we report on the extension of this technique to small ion Coulomb crystals made of two or three  $^{40}$Ca$^+$ ions. Efficient cooling of the axial motion is achieved outside the Lamb-Dicke regime on a two-ion string along the magnetic field axis as well as on two- and three-ion planar crystals. Complex sideband cooling sequences are required in order to cool both axial degrees of freedom simultaneously. We measure a mean excitation after cooling of $\bar n_\text{COM}=0.30(4)$ for the centre of mass mode and $\bar n_\text{B}=0.07(3)$ for the breathing mode of the two-ion string with corresponding heating rates of {11(2)}\,s$^{-1}$ and \SI{1(1)}{\per\second} at a trap frequency of \SI{162}{\kilo\hertz}. The ground state occupation of the axial modes is above 75\% for the two-ion planar crystal and the associated heating rates 0.8(5) s$^{-1}$ at a trap frequency of \SI{355}{\kilo\hertz}.

\begin{keywords}Penning trap; trapped ions;
sideband cooling; ion Coulomb crystals
\end{keywords}\bigskip

\end{abstract}
    
    
\maketitle
\section{Introduction} 

Ion Coulomb crystals (ICCs) consisting of cold, trapped atomic ions are a widely used and highly versatile experimental platform \cite{Thompson2015}. The level of control achievable with ICCs makes them a suitable choice for many applications in atomic and molecular physics, including quantum computation and simulation~\cite{Britton2012,Debnath2016,Monz2016}, cavity QED \cite{Herskind2009}, atomic clocks~\cite{Chou2010} and precision measurements~\cite{Leanhardt2011}. Penning traps use a combination of static electric and magnetic fields to confine charged particles. Such traps are often used in experiments where large magnetic fields are required, 
for example in quantum simulation \cite{Britton2012}, non-neutral plasma physics \cite{Bollinger2003}, precision measurements of
masses and magnetic moments \cite{Blaum2010}, and for experiments on particle beamlines (where their large, open electrode structures and large trap depths are an advantage)
\cite{Smorra2015,Murbock2016}.

In quantum information experiments, the common modes of motion of ICCs provide a convenient mechanism for transmitting information between ions~\cite{CZ95}. Radio-frequency, linear Paul traps are prevalent in this kind of work, typically using one-dimensional crystals that align along the radio frequency null of the trap~\cite{Lanyon2013}.  Such a system with three ions in a line has been used to simulate the frustration that occurs when three spins are arranged in a triangle configuration \cite{Kim2010}. In contrast, planar ICCs in a Penning trap naturally form in a triangular lattice, which provides a suitable platform for quantum simulation of frustrated systems~\cite{Bohnet2016}, without the need for highly specialised trap designs~\cite{Yoshimura2015,Mielenz2016}.

In order to conduct many of these types of experiments, the ions must be in the Lamb-Dicke regime, where the amplitude of the motional mode is much less than the wavelength of the interacting light.  This often requires the ions to be cooled below the Doppler limit, most commonly achieved in ion traps using the resolved-sideband technique. 
Here we present results on resolved-sideband laser cooling of the axial motion of two- and three-ion $^{40}$Ca$^+$ ICCs in a Penning trap. 

We dedicate this article to the memory of our colleague Professor Danny Segal whose inspiration and insight led to many of the ideas  that are described here.  Over many years Danny designed and built the apparatus and developed the techniques for working with ground-state cooled ions in the Penning trap and he anticipated the efficient sideband cooling and long coherence times that we have now demonstrated for small ion Coulomb crystals.  Sadly he was only partially able to see his ideas come to fruition as he passed away in September 2015 before this work had been completed.

\section{Background} 
To confine charged particles in all directions of space, a Penning trap uses a strong magnetic field aligned with the trap axis ($\vec{B}=B\hat{z}$) and a static cylindrically-symmetric quadrupolar electric potential. Expressed in cylindrical coordinates, the potential has the form $\phi=A(2z^2-\rho^2)$ where $A$ is a constant depending on the voltage applied on the electrodes of the trap and the geometry of the trap. In the axial direction (along $\hat{z}$), a single particle will undergo a simple harmonic motion with a frequency $\nu_z=\sqrt{4qA/M}/2\pi$, where $q$ is the charge of the particle and $M$ is its mass.  Typically this frequency lies in the range 150 to 400\,kHz in our trap. Confinement in the radial plane is ensured by the magnetic field, resulting in stable orbits about the axial direction. The motion in this plane is characterised by the modified cyclotron and the magnetron modes whose  frequencies $\nu_+$ and $\nu_-$ are given by $\nu_{\pm}=\nu_c/2\pm\sqrt{\nu_c^2/4-\nu_z^2/2}$, where the cyclotron frequency, $\nu_c$, is given by $\nu_c=qB/(2\pi M)$.  In our trap, the cyclotron frequency is around 715\,kHz.
    
Multiple ions in a trap exert a repulsive Coulomb force on one another which counteracts the trapping force. If these ions are sufficiently cold, they form a structure known as an ion Coulomb crystal, where the individual ions exhibit small amplitude motion around their equilibrium positions \cite{Thompson2015}. An ICC consisting of $N$ particles has a total of 3$N$ of these collective modes of motion. Due to the cylindrical symmetry of the Penning trap potentials and the presence of the axial magnetic field, one of the radial modes corresponds to  a bulk rotation of the crystal and for stable radial confinement the frequency of this rotation, $\nu_r$, must be between $\nu_-$ and $\nu_+$.
    
\begin{figure}[h] 
\begin{center}
\includegraphics[width=0.9\columnwidth]{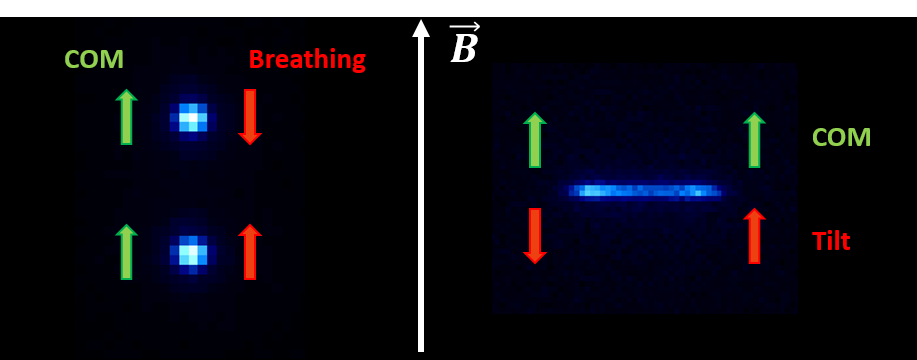}
\caption{EMCCD images of two ions in a string (left) and in a planar crystal (right) as seen from the radial plane. The planar crystal appears as a blurred line due to its rotation about the magnetic field axis.    }
\label{ModesDiagram}
\end{center}
\end{figure}

The effective radial trapping strength in the crystal's frame of reference is described by the frequency $\nu_{\text{eff}}$, which can be expressed as a function of $\nu_r$  by
\begin{equation}
\nu_{\text{eff}}=\sqrt{\nu_r (\nu_c - \nu_r) -\tfrac{1}{2}\nu_z^2}\ ,
\label{freqEff}
\end{equation}
which is maximum when the rotation frequency is half the cyclotron frequency, i.e., $\nu_r = \frac{1}{2} \nu_c$. For a given rotation frequency, an increase in the axial trapping strength (i.e., an increase in the endcap voltage) leads to a decrease in the effective radial trapping strength and a change in the shape of the ICC~\cite{Mavadia2013}. At low axial frequency the ions can be made to line up along the magnetic field axis (an `axial string'), and by increasing the axial frequency the ions can be made to lie in the radial plane (a `planar crystal'). These two configurations are of particular interest, as the axial and radial modes are separable and the simpler axial modes can be addressed independently.  Only the axial modes of ICCs are discussed in this paper.

Figure \ref{ModesDiagram} shows the modes of two-ion ICCs in the string and planar configurations. A two-ion string has an axial centre of mass (COM) mode at a frequency $\nu_z$, and a breathing mode (where the ions move in opposite directions) at $\sqrt{3}\nu_z$. For planar crystals there is the COM mode at $\nu_z$ and a `tilt' mode at $\nu_{\text{tilt}}=\sqrt{\nu_z^2- \nu_{\text{eff}}^2}$. In the tilt mode the ions have equal and opposite axial displacements.  For a three-ion planar ICC there are two degenerate tilt modes at $\nu_{\text{tilt}}$. The degeneracy of these two modes could be lifted by the application of a rotating wall drive, which breaks the cylindrical symmetry of the system~\cite{Hasegawa2005}. 
    
The interaction Hamiltonian for the axial motion of a two-level, multi-ion crystal uniformly illuminated by a plane electromagnetic wave of frequency $\omega_L/2\pi$ is given by

\begin{equation}
 H_I= \frac{\hbar}{2} \sum_{j=1}^{N} \Omega_0 (\sigma_j^++\sigma_j^-) e^{-i \sum_k \eta_k^{(j)} ( a_k^\dagger + a_k) } e^{i\omega_Lt}+ \text{h.c.}
 \label{hamiltonian}
\end{equation}
where $\Omega_0$ is the Rabi frequency in the absence of ion motion, $a_k^\dagger$ and $a_k$ are the phonon raising and lowering operators for the $k$-th mode and $\eta_k^{(j)}$ is the Lamb-Dicke parameter for the $j$-th ion in the $k$-th mode, which takes into account the different amplitude of motion for different ions in a particular mode. 

\begin{figure}[h] 
\begin{center}
\includegraphics[width=0.8\columnwidth]{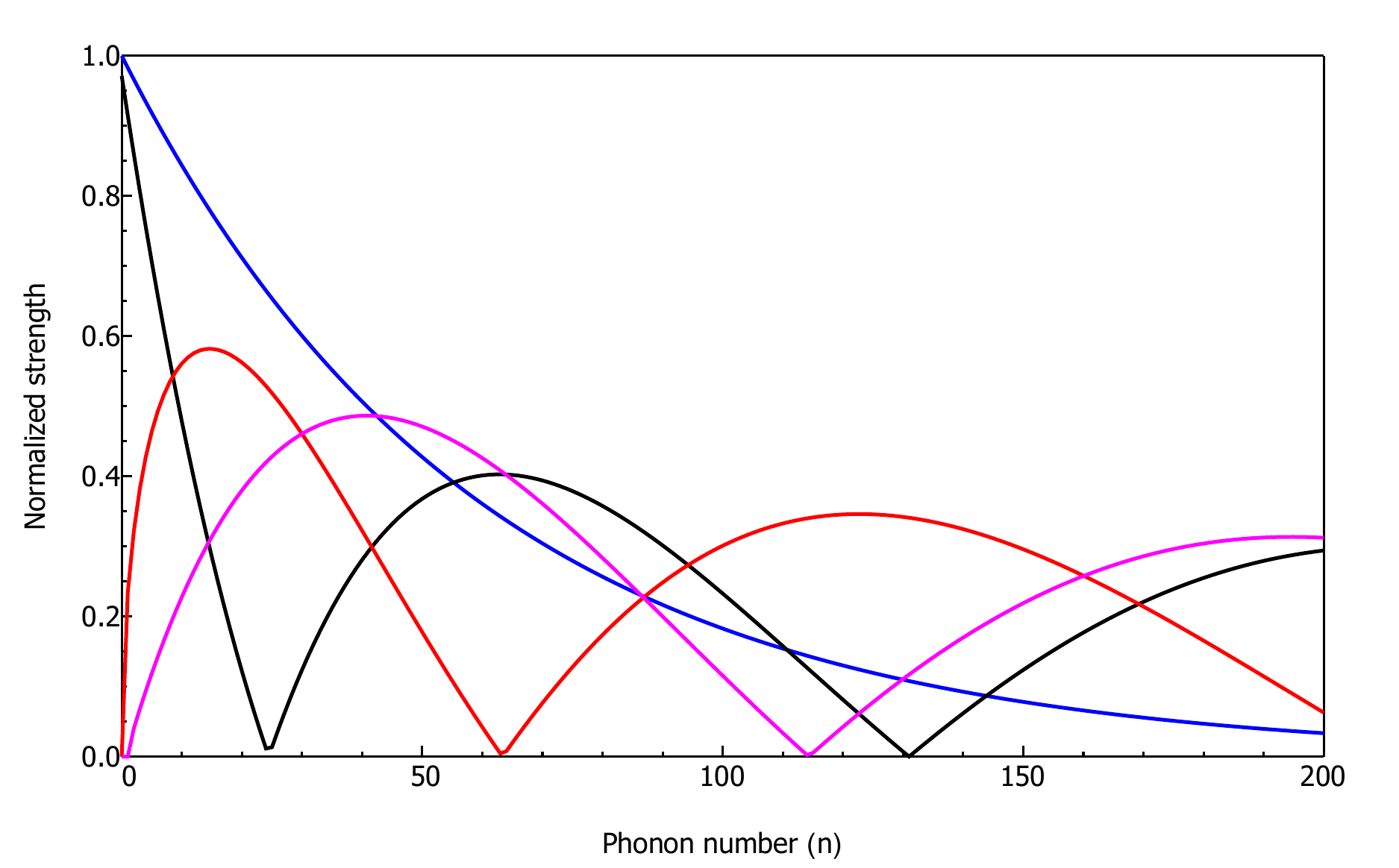}
\caption{Normalised coupling strength at the carrier (black), first red sideband (red) and second red sideband (magenta) for an axial frequency of \SI{162}{\kilo\hertz} (for which $\eta_0=0.24$). The blue line shows the population distribution $P(n)$ in a thermal state at the Doppler limit ($\bar{n}=58$), normalised to $P(0)=1$. }
\label{sidebands}
\end{center}
\end{figure}

Let us first consider what happens in the case of a single ion with oscillation frequency $\nu_z$. In this case the Lamb-Dicke parameter is given by
\begin{equation}
\eta_0=\frac{1}{\lambda}\sqrt{\frac{h}{2M\nu_z}}\ ,
\end{equation}
where $\lambda$ is the wavelength and $h$ the Planck constant. The Lamb-Dicke regime is defined by the condition that $\eta_0\sqrt{2n+1}\ll1$, where $n$ is the quantum number of the axial motion. 
The axial motion gives rise to a set of sidebands around the carrier frequency spaced at the corresponding oscillation frequency ($\nu_z$), where the sideband at frequency $m\nu_z$ corresponds to transitions from state $n$ to state $n'=n+m$. The relative Rabi frequencies associated with each sideband (and the carrier, with $m=0$) as a function of phonon number $n$ are proportional to associated Laguerre polynomials \cite{Leibfried2003}. This is shown in Figure~\ref{sidebands} for a trapping frequency of $\nu_z=\SI{162}{\kilo\hertz}$, 
where it can be seen that there are values of $n$ at which the amplitude of a sideband becomes very close to zero. We will refer to these points as `minima'.  These points give rise to what are in effect dark states at particular values of $n$, where there is very little interaction between the ion and the laser when it is tuned to that sideband.  In particular, in this case, there is a minimum for the carrier around $n=24$ and for the first sideband around $n=63$. These minima move to lower values of $n$ as $\eta_0$ rises. The figure also shows the thermal distribution at the Doppler limit expressed in terms of the relative occupation probability $P(n)$ for each Fock state $n$ (normalised to unity for $n=0$).

At low trapping frequencies, and therefore at large Lamb-Dicke parameters, an ion after Doppler cooling will be left in a thermal distribution with significant parts of the population at phonon numbers higher than the lowest minimum of the first red sideband. The standard sideband cooling technique, which consists of exciting only the first red sideband, will leave this part of the population `trapped' at the minimum, where only a very small rate of cooling can occur, and a high ground state occupation probability will therefore not be achieved. However, because the minima are located at a different  phonon number for each sideband,  sideband cooling well outside the Lamb-Dicke regime remains possible but requires different red sidebands (typically the first and second alternately) to be addressed~\cite{Poulsen2012}.  This is the procedure that we adopted in the sideband cooling of a single ion reported in Ref. \cite{Goodwin2016}.

We now consider the extension to the case of two ions.  In this situation, the Lamb-Dicke parameters for both ions are always equal to each other.  The centre of mass mode of a two-ion crystal always has a Lamb-Dicke parameter of $\eta_{\text{COM}}=\eta_0/\sqrt{2}$ because the mass of the crystal is twice that of a single ion. To a good approximation the Lamb-Dicke parameter for the tilt mode of a radial crystal also has this value in our experiments.  However, the breathing mode of the axial crystal has $\eta_{\text{B}}=\eta_0/\sqrt{2\sqrt{3}}$ \cite{Morigi1999}. 

The Hamiltonian in Equation \ref{hamiltonian} gives rise to a spectrum of sidebands at the harmonics of each mode and at the frequencies of all the corresponding intermodulation products. Outside the Lamb-Dicke regime, where sidebands beyond first order are significant, the spectrum can become very complicated. The Rabi frequency for a sideband transition between initial phonon states $(n_1,n_2)$ and final phonon states $(n'_1,n'_2)$ is given by
\begin{equation}
 \Omega_{n_1',n_2',n_1,n_2} = \Omega_0 \prod_{k=1}^2\left[\sqrt{\frac{\min{\{n_k,n'_k\}}!}{\max{\{n_k,n'_k\}}!}} \eta_k^{|n'_k-n_k|} e^{-\eta_k^2/2} L_{\min{\{n_k,n'_k\}}}^{|n'_k-n_k|}(\eta_k^2)\right]\label{Rabi}
 \end{equation}
where $L$ is a Laguerre polynomial. Here $k=1$ refers to the COM mode and $k=2$ refers to the breathing mode. This equation shows that 
the strength of any sideband described by $\Omega_{n_1',n_2',n_1,n_2}$ depends on all four quantum numbers. In particular, the strength of any sideband of one motion is proportional to the strength of the \emph{carrier} of the other motion.

\begin{figure}[h] 
\begin{center}
\includegraphics[width=0.8\columnwidth]{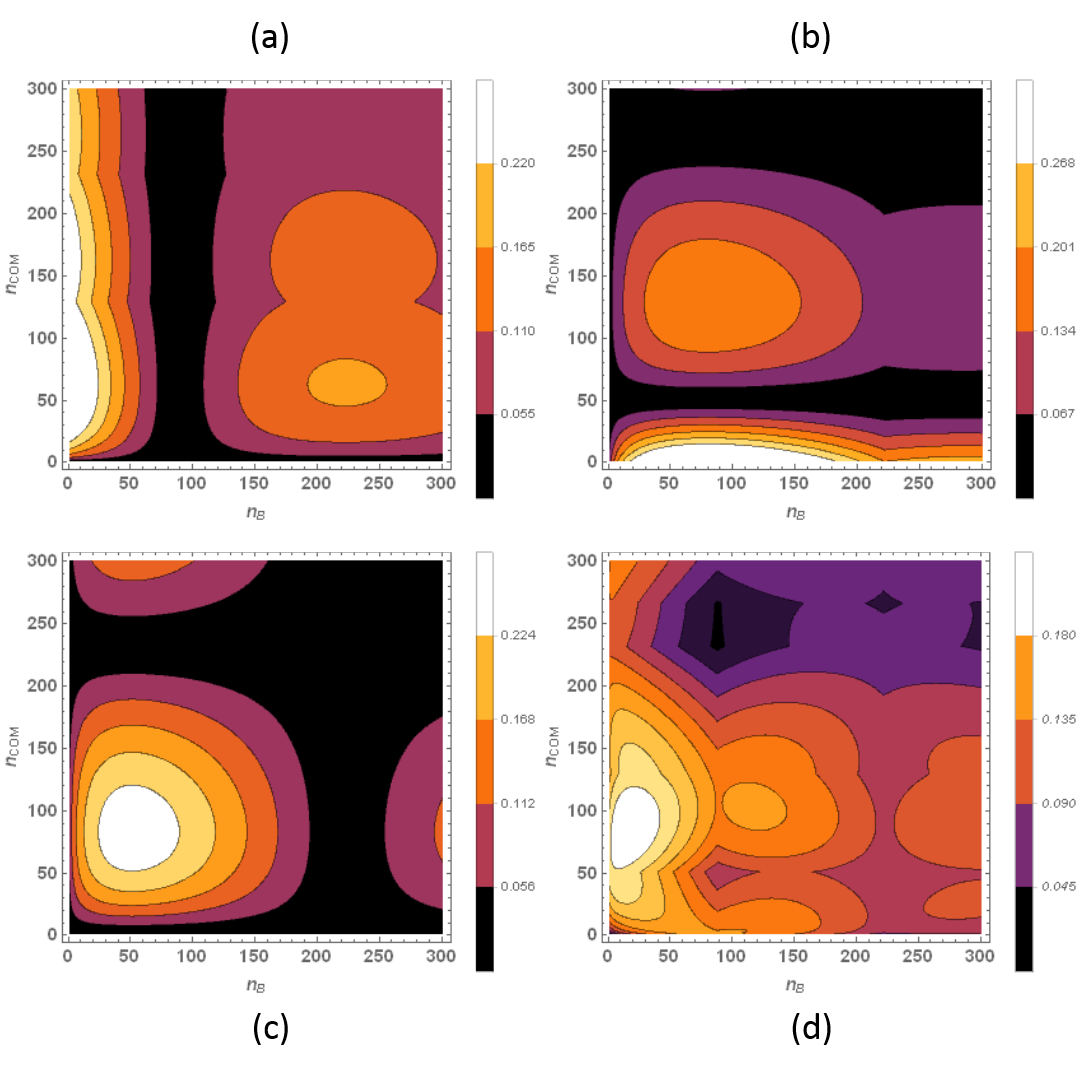}
\caption{Rabi frequency of various sidebands for a two-ion axial string at $\nu_z=\SI{162}{\kilo\hertz}$ when cooling is carried out (a) on the first, second and third red COM sidebands, (b) on the first and second red breathing mode sidebands, (c) on the  first breathing mode red sideband of the second red COM sideband and (d) an average of plots (a), (b) and (c). The Lamb-Dicke parameters at this frequency have the values 0.17 for the COM mode and 0.13 for the breathing mode. }
\label{TwoIonStrength}
\end{center}
\end{figure}



As a result, the simultaneous cooling of the two motional modes of a two-ion axial string outside the Lamb-Dicke regime is significantly more difficult than for a single ion. We will discuss here  the case of a two-ion string, which is more challenging to cool than the planar structure, but the reasoning would be similar for the latter. As in the case of a single ion, sideband cooling of two ions becomes very slow at some phonon numbers where the coupling strength of the first red sideband of either motional mode approaches zero. This problem can be addressed by cooling on higher-order sidebands in a similar fashion to the single-ion case. Additionally, sideband cooling on a given mode will also become extremely slow  for phonon numbers where the carrier strength of the \emph{other} mode is at a minimum. 
However there are some combinations of quantum numbers that give population trapping where the carrier strength for \emph{both} modes is at a minimum. Such `dark' regions can be identified in a two-dimensional phonon space where each dimension corresponds to one of the motional modes. 

These regions are represented in Figure \ref{TwoIonStrength}. The top left quadrant (a) shows the collective strength of the first three red sidebands of the centre of mass  mode, representative of the cooling rate for a sequence that alternates between these sidebands. The darker regions correspond to phonon numbers where the carrier strength of the breathing mode is at a minimum. A similar plot for the breathing mode (b), where only the two first red sidebands need to be considered, shows an analogous situation. It can be seen, by overlapping the two plots, that the brighter areas of one eliminate most of the dark regions of the other. This means that by alternating cooling between the two modes, most of the population trapping can be avoided. However, there remains a dark area at the intersection of the minima of the carriers of the two modes. This region is centred around $n_\text{COM}=49$ and $n_\text{B}=86$.  The population trapped at this intersection can only be brought to the ground state by including an intermodulation product sideband in the cooling sequence, that is, a sideband that changes the phonon number for both modes at the same time. A plot of the strength of the second red COM sideband of the first red breathing mode sideband is shown on Figure \ref{TwoIonStrength}(c), which features a bright region at the dark crossing area of (a) and (b). This sideband therefore cools population that has accumulated in the dark region efficiently to lower phonon numbers, allowing further cooling on the other sidebands to continue.  On the last quadrant (d) where the three other plots are overlapped, no dark region remains, thus preventing population trapping over the complete cooling sequence. More detailed simulations we have carried out allow us to optimise the cooling sequence to obtain the highest ground-state population in both modes~\cite{Joshi2017}. 

Once the ions are cooled to the Lamb-Dicke regime, the interaction Hamiltonian (Equation~\ref{hamiltonian}) can be greatly simplified in the interaction picture by only considering terms to low orders in $\eta_k$. With this approximation, the Hamiltonian for two ions can be written as:

\begin{eqnarray}
H_I&=& \frac{ \Omega_0\hbar}{2} \sum_{j=1}^{2}  e^{-i\delta_j t
}\ket{\downarrow_j}\bra{\uparrow_j} \times \nonumber \\ 
&& [(1-\eta_1^2(n_1+\tfrac{1}{2})-\eta_2^2(n_2+\tfrac{1}{2}))\ket{n_1,n_2}\bra{n_1,n_2}+ \nonumber \\ 
&& i\eta_1(e^{i\nu_1t}\sqrt{n_1}\ket{n_1-1,n_2}\bra{n_1,n_2}+e^{-i\nu_1t}\sqrt{n_1+1}\ket{n_1+1,n_2}\bra{n_1,n_2}) + \nonumber \\ 
&& i\eta_2(e^{i\nu_2t}\sqrt{n_2}\ket{n_1,n_2-1}\bra{n_1,n_2}+e^{-i\nu_2t}\sqrt{n_2+1}\ket{n_1,n_2+1}\bra{n_1,n_2})]
+ \text{h.c.}
\label{TwoIonHam}
\end{eqnarray}
In this equation the second line describes the carrier, the third line the red and blue sidebands of mode 1 and the fourth line the red and blue sidebands of mode 2.  
Note that the detuning $\delta_j$ is not necessarily constant across all ions $j$, despite using a single laser to excite the whole crystal. This becomes important in the case of the axial string of two ions, where their carrier frequencies differ by approximately \SI{2}{\kilo\hertz} in our experiment due to the axial magnetic field inhomogeneity.

\section{Methods}  
The apparatus used to perform this experiment is largely the same as that previously used to sideband cool single ions to the quantum ground state~\cite{Goodwin2016} and for observing and controlling the spatial configurations of ICCs~\cite{Mavadia2013}. 
The Penning trap consists of stacked cylindrical electrodes with an internal diameter of \SI{21.6}{\milli\meter}, housed in a vacuum system and inserted into the bore of superconducting magnet producing a magnetic field strength of approximately \SI{1.865}{\tesla} at the field centre. DC voltages are applied to two of the electrodes which act as end-caps, and `axialisation drive' voltages applied to a segmented ring electrode produce an oscillating quadrupole potential to aid Doppler cooling of the radial motion~\cite{Powell2002a}. A cloud of $^{40}$Ca$^+$ ions is loaded into the trap via three-photon non-resonant photoionisation of an atomic beam using a frequency-doubled pulsed Nd:YAG laser. The ions are then cooled to close to the Doppler limit by addressing $\textrm{S}_{\nicefrac{1}{2}}\leftrightarrow\textrm{P}_{\nicefrac{1}{2}}$ electric dipole transitions at \SI{397}{\nano\metre}, as well as relevant repumping transitions~\cite{Mavadia2014}. Doppler cooling beams propagate both along and perpendicular to the direction of the magnetic field to ensure that all of the motional modes of the crystal are cooled~\cite{Itano1982,Mavadia2013}. The radial cooling beam is offset from the trap centre, resulting in an intensity gradient across the crystal and giving coarse control of the crystal rotation frequency~\cite{Itano1988,Asprusten2014}. Note that good control of the radial beam position and intensity is paramount for the stability of the crystal. Unlike for a single ion, the axialisation drive has to be fairly strong for ICCs to maintain the crystal configuration and achieve efficient Doppler cooling. The results presented here for ICCs were obtained with \SIrange{2}{4}{\volt} of axialisation while a \SIrange{50}{200}{\milli\volt} signal is typically enough to efficiently cool the radial motion of a single ion in our trap.
    
The crystals are sideband cooled using broadly the same method used for single ions~\cite{Goodwin2016}, addressing specific motional sidebands of the $\textrm{S}_{\nicefrac{1}{2}}(m_j=-\frac{1}{2})\leftrightarrow\textrm{D}_{\nicefrac{5}{2}}(m_j=-\frac{3}{2})$ transition using a narrow-linewidth laser at \SI{729}{\nano\metre} propagating along the trap axis. The frequency of this laser is controlled using an acousto-optic modulator and a direct digital synthesis (DDS) frequency generation system, 
which permits sequential addressing of up to seven sidebands during each cooling sequence. An additional `quench' laser at \SI{854}{\nano\metre} is applied to artificially increase the effective decay rate from the $\textrm{D}_{\nicefrac{5}{2}}$ state and therefore increase the cooling rate~\cite{Marzoli1994}.
    
After sideband cooling the ions are prepared in the $\textrm{S}_{\nicefrac{1}{2}}(m_j=-\frac{1}{2})$ state by optical pumping and then probed on the $\textrm{S}_{\nicefrac{1}{2}}(m_j=-\frac{1}{2})\leftrightarrow\textrm{D}_{\nicefrac{5}{2}}(m_j=-\frac{3}{2})$ spectroscopy transition (the same as that used for sideband cooling). We then perform a projective measurement of the state of the ions by turning on the Doppler cooling lasers and collecting fluorescence from the ions using either an EMCCD camera or photomultiplier tubes \cite{Mavadia2014}. The state of each of the individual ions in an axial string can be determined using the EMCCD camera, which spatially resolves the ions. For planar ICCs the bulk rotation of the crystal prevents resolution of different ions using side-on imaging (as seen on Figure \ref{ModesDiagram}) and the total number of photons received must therefore be used to distinguish between different numbers of bright ions. This cooling, probe and detection cycle is repeated many times (typically 200) at a particular probe laser frequency to measure the excitation probability, before stepping the frequency to obtain a full sideband spectrum.

\section{Results and discussion}  
 Before we attempt to sideband cool or perform spectroscopy on an ICC, we first observe it on an EMCCD camera during continuous Doppler cooling and adjust a combination of the endcap voltage, radial beam offset and axialisation drive amplitude until the crystal remains stable in the desired configuration. For an axial string to form we require that the axial trapping strength is sufficiently weaker than the radial trapping strength (with the converse being true for planar crystals). However, the effective radial trapping strength depends on the rotation frequency of the crystal.  It is maximum when the crystal rotates at half the cyclotron frequency.  This defines a critical trapping voltage above which the two-ion crystal will always lie in the radial plane.  However, below this voltage the crystal may still lie in the radial plane because the radial trapping strength (described by $\nu_{\text{eff}}$) is reduced for different rotation frequencies.  Indeed, there is always some rotation frequency for which $\nu_{\text{eff}}<\nu_z$ so the crystal cannot be forced into the axial string configuration solely by relaxing the axial potential, without some control over the crystal rotation frequency. We find that the end-cap voltage must be tuned significantly below the calculated critical value in order to achieve axial strings that remain stable for the duration of an experiment.

\subsection{Two-ion axial string}  

\begin{figure}[h]  
\begin{center}
\includegraphics[width=0.8\columnwidth]{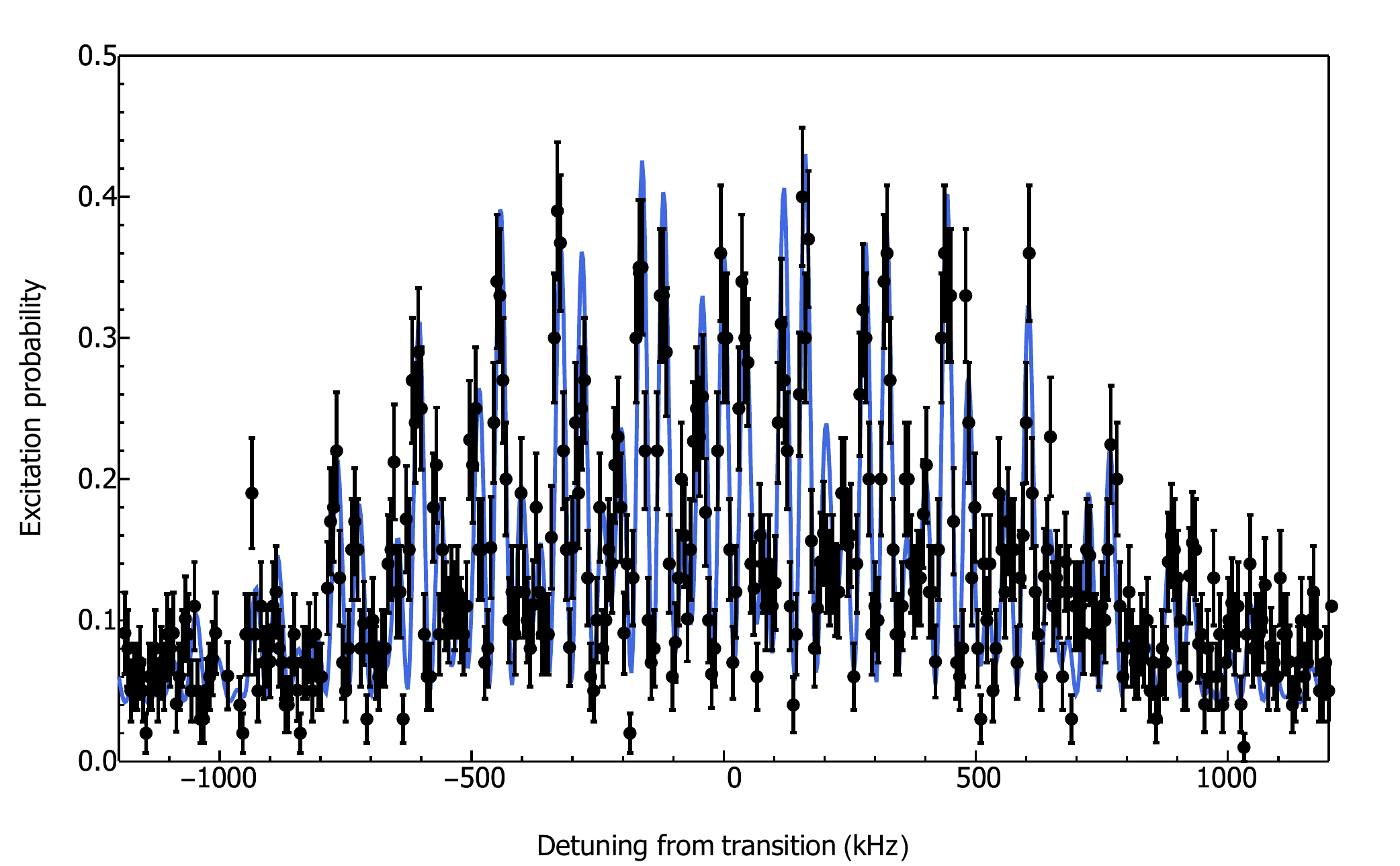}
\caption{Spectrum of the excitation probability of one ion in a two-ion string after Doppler cooling at a trapping frequency $\nu_z=\SI{162}{\kilo\hertz}$. A simulated spectrum (blue line) is superimposed on the data points (in black).  {  The simulation corresponds to average phonon numbers $\bar{n}_\text{COM}=87$ and $\bar{n}_\text{B}=51$. } See the text for details of how the simulation is produced.} 
\label{2ionAxialDoppler}
\end{center}
\end{figure}

Prior to sideband cooling it is essential to pre-cool the ions to their Doppler limit. Figure \ref{2ionAxialDoppler} shows a spectrum obtained after Doppler cooling of a two-ion string at a trapping frequency (COM) $\nu_z=\SI{162}{\kilo\hertz}$. At this frequency, the Lamb-Dicke parameter for both modes is fairly high: 0.17 for the COM mode and 0.13 for the breathing mode, consistent with the many sidebands visible. Due to the complexity of the spectrum, a fit to the data was not attempted. Instead, a spectrum was simulated and the parameters adjusted so that the simulated curve approaches the data points. 
In this simulated spectrum, the average phonon number is set to be $\bar{n}_\text{COM}$ = 87 and $\bar{n}_\text{B}$ = 51 (these values are approximately 1.5 times the Doppler limit). The occupancy of each phonon state $\ket{n_\text{COM}}\otimes\ket{n_\text{B}}$ is calculated and the excitation strength is evaluated numerically from the two-dimensional Rabi strength in Equation \ref{Rabi}. The Rabi frequency and the pulse length are fixed to the values that are used to perform the experiment. The contribution due to each phonon state and  motional sideband is summed at a particular laser detuning and thus a spectrum is generated which matches the experimental data reasonably well.   
Using the parameters of the simulation, and assuming a thermal distribution, the probability for $n_\text{COM}$ to remain above the first minimum of the red sideband after Doppler cooling is about 24\,\% while it is less than 1.4\,\% for the breathing mode.  
    
\begin{figure}  
\begin{center}
\includegraphics[width=0.8\columnwidth]{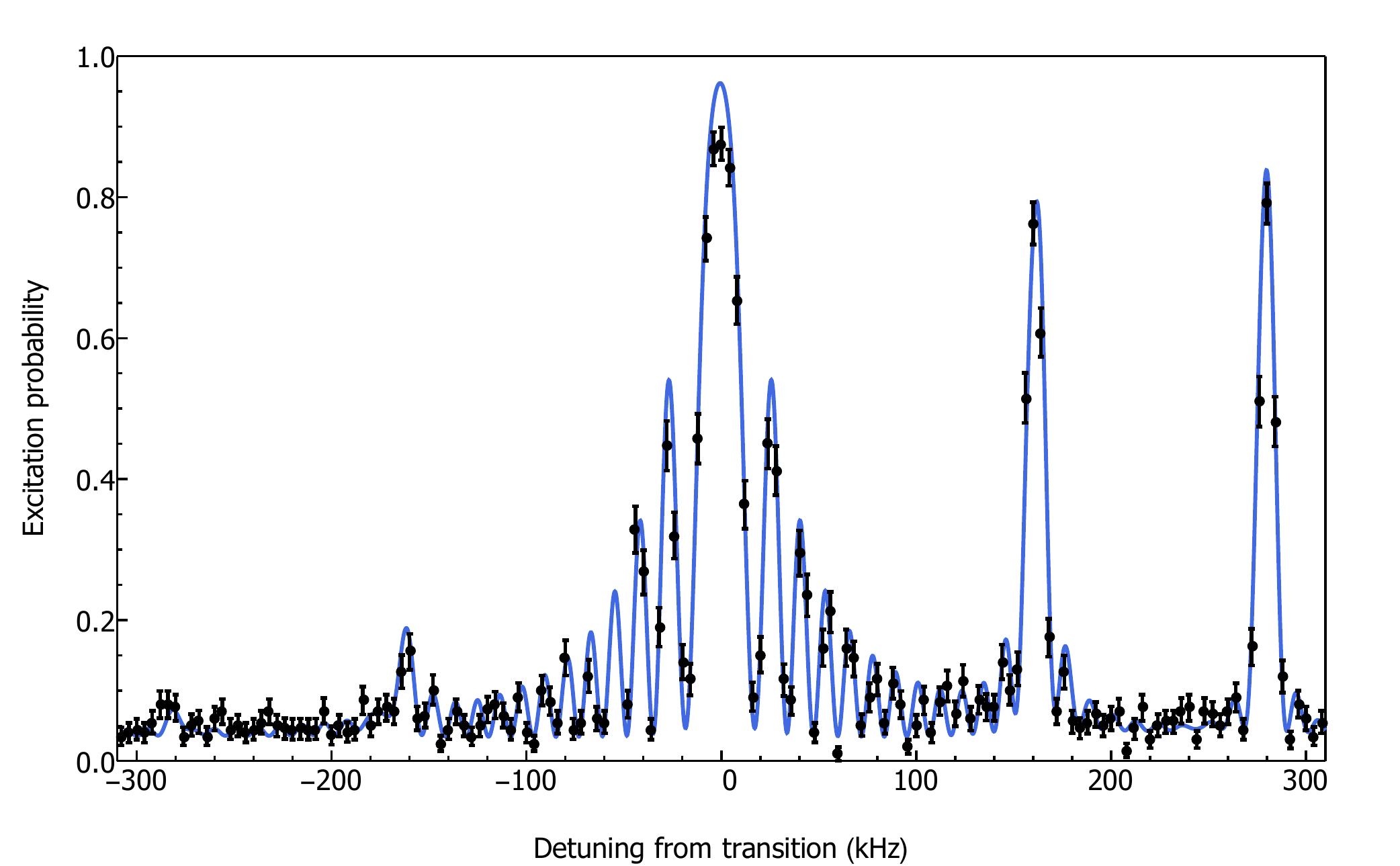}
\caption{Spectrum of the excitation probability of one ion in a sideband-cooled two-ion axial crystal at an axial frequency of 162\,kHz, showing a fit to the carrier and the first-order sidebands of both motions (see text for details).  
}
\label{2ionAxialSBC}
\end{center}
\end{figure}

To sideband cool both modes simultaneously and avoid population trapping, several sidebands are addressed sequentially with laser pulses of a few hundreds of microseconds. Table~\ref{twoIonCoolSeq} shows the pulse sequence used to sideband cool the two-ion string. It is the result of both theoretical studies and empirical adjustments to obtain the lowest final temperature. Figure~\ref{2ionAxialSBC} shows the spectrum after sideband cooling. It can be seen that the sideband structure simplifies greatly compared to the Doppler spectrum.  The strongest visible components  are now just the carrier and the first blue sidebands of the COM and breathing modes. There are small peaks present at the positions of the first red sidebands of the two motions.  
The large asymmetry between the blue and red sidebands indicates a high ground state occupation probability. 
In order to extract the average phonon number, we assume a thermal distribution among the  lowest five states of each mode (i.e., $n=0$ to 4).  Numerical solutions of the Schr\"{o}dinger equation are constructed using the Hamiltonian (Equation \ref{TwoIonHam}) for the carrier and first red and blue sidebands of both modes. This numerical solution correctly takes into account the entanglement created between the two ions, based on the method discussed in \cite{Homethesis}.  Using the spatial information from the camera, the fluorescence data of   each ion was recorded separately.  The two spectra were fitted independently to the excitation probabilities of the ions calculated using our model, taking into account the small frequency difference of the two carriers due to the magnetic field inhomogeneity. 
The extracted parameters were then averaged to obtain the final mean phonon numbers  $\bar{n}_\text{COM}=0.30(4)$ for the COM mode and $\bar{n}_\text{B}=0.07(3)$ for the breathing mode. We measured the heating rates by inserting delay periods after sideband cooling but before probing the ions.  We obtain heating rates of {11(2)}\,s$^{-1}$ for the COM and \SI{1(1)}{\per\second} for the breathing mode.

\begin{table}
\begin{center}
\tbl{Sequence of red sidebands addressed to sideband cool a two ion axial string (COM = centre of mass mode; B = breathing mode).}
{\begin{tabular}{l  c  c}
\toprule
Sideband & Pulse time ($\upmu$s) & Sequence repeats\\
\hline
\hline
2\textsuperscript{nd} COM & 500 &\\
2\textsuperscript{nd} B & 500&\\
3\textsuperscript{rd} COM & 300& 15\\
1\textsuperscript{st} B & 200&\\
2\textsuperscript{nd} COM 1\textsuperscript{st} B & 500&\\
\hline
1\textsuperscript{st} B & 200&\\
2\textsuperscript{nd} COM & 500 & 2\\
1\textsuperscript{st} COM & 500 &\\
2\textsuperscript{nd} COM 1\textsuperscript{st} B & 500&\\
\hline
2\textsuperscript{nd} COM 1\textsuperscript{st} B & 500&\\
1\textsuperscript{st} COM & 2000 & 1\\
1\textsuperscript{st} B & 500&\\
\botrule
\end{tabular}}
\label{twoIonCoolSeq}
\end{center}
\end{table}
    
\subsection{Planar crystals} 
    
Working with planar crystals is a somewhat simpler proposition as the axial frequency is necessarily higher than for an axial crystal having the same number of ions. This results in a lower value of $\eta$ and therefore there are fewer visible sidebands after cooling to the Doppler limit. In contrast to axial crystals, there are trapping voltages at which the crystal is always stable in the radial plane. 
One of the main limiting factors is our ability to precisely control the rotation frequency using only the torque provided by an offset radial cooling beam. Greater control could be achieved using a quadrupolar or hexapolar rotating wall drive~\cite{Hasegawa2005,Khan2015}.
    
\begin{figure} 
\begin{center}
\includegraphics[width=0.8\columnwidth]{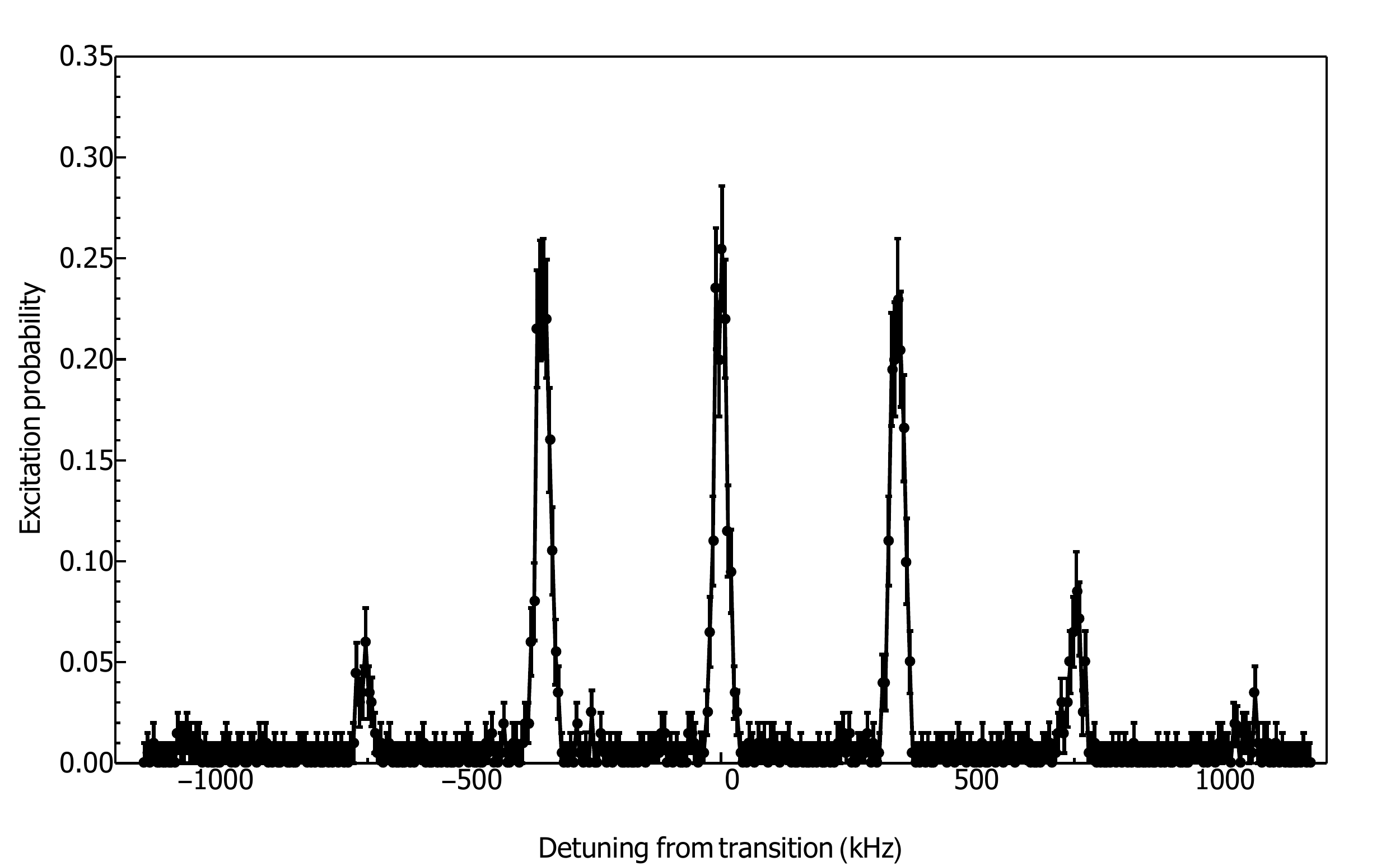}
\caption{Spectrum of a Doppler-cooled, two-ion planar crystal at a trapping frequency of $\nu_z=\SI{353}{\kilo\hertz}$.  The ordinate represents the probability of excitation of both ions simultaneously.  The solid line is included as a guide to the eye and does not correspond to a fit to the data.}
\label{2ionPlanarDoppler}     
\end{center}
\end{figure}

Figure~\ref{2ionPlanarDoppler} shows the spectrum of a two-ion planar ICC after Doppler cooling. This spectrum is taken at a trapping frequency of $\nu_z=\SI{353}{\kilo\hertz}$ (corresponding to $\eta_\text{COM}=0.115$). The rotation frequency of the crystal is such that the tilt mode is very close in frequency to the COM mode and their corresponding sidebands are not distinguishable on the spectrum. 

    
Owing to the relatively small Lamb-Dicke parameter and the near degeneracy of the COM and tilt modes, the pulse sequence for sideband cooling of the planar crystal is much simpler than for the string. Pulses are only applied on the COM red sidebands -- the tilt mode is excited off-resonantly -- and a maximum of just two sidebands (first and second red) need to be addressed to prevent any significant population trapping.


    
\begin{figure} 
\begin{center}
\includegraphics[width=0.8\columnwidth]{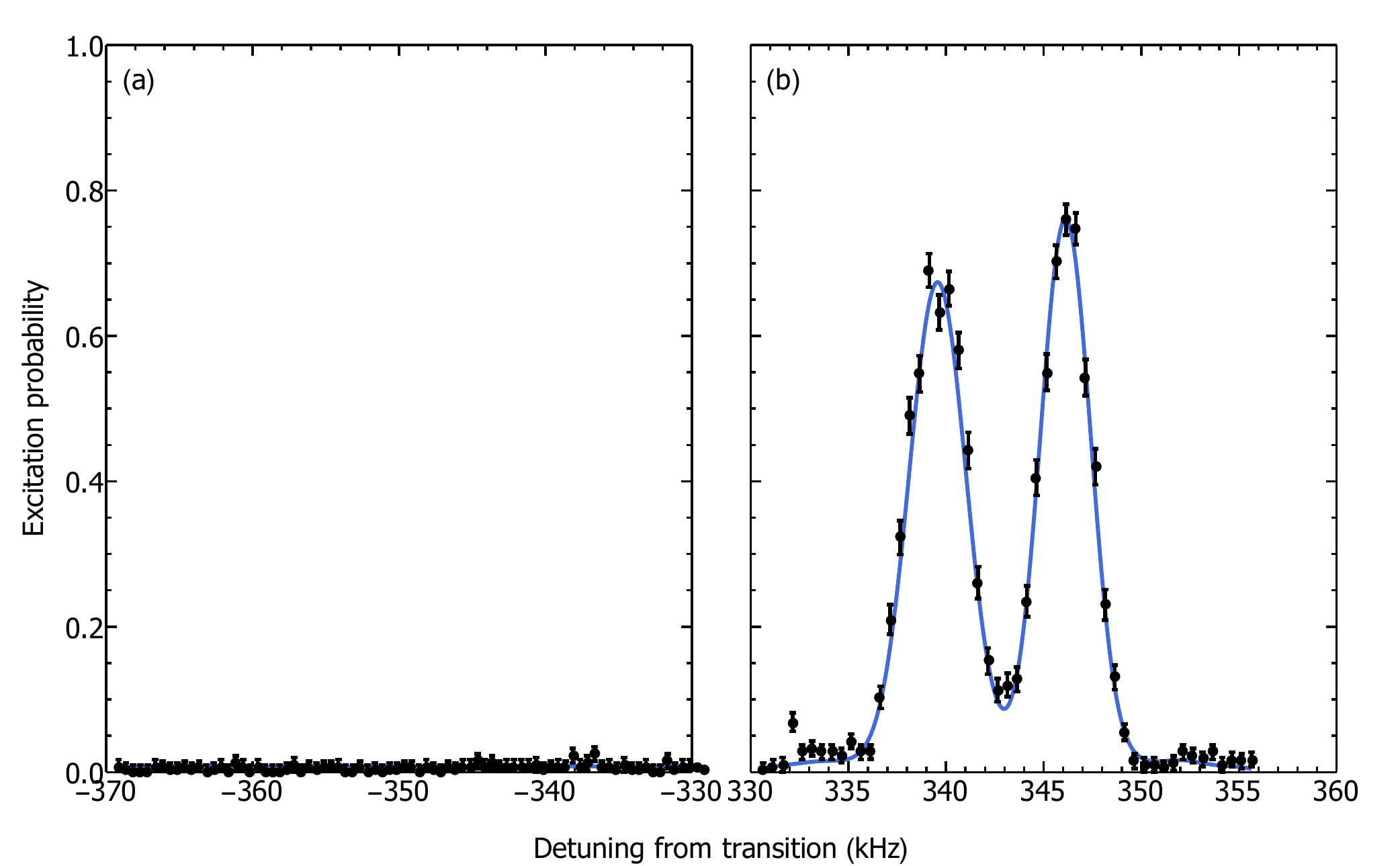}
\caption{Spectrum of a two-ion planar crystal after \SI{10}{\milli\second} of sideband cooling at a trapping frequency $\nu_z=\SI{346}{\kilo\hertz}$ showing the suppressed first red sidebands (a) and first blue sidebands (b). The peak at a higher frequency corresponds to the COM mode. The blue line is a fit to data points (see text). The fit gives a final phonon number for both modes of $\bar{n} < 0.25$. The ordinate represents the probability of excitation of both ions simultaneously.  Note that there are artifacts corresponding to rotational sidebands present at other frequencies in the spectrum, but they are not showed in this windowed spectrum.}
\label{2ionPlanarSBC}
\end{center}
\end{figure}

Figure~\ref{2ionPlanarSBC} shows the spectrum obtained after sideband cooling of a two-ion planar ICC at $\nu_z=\SI{346}{\kilo\hertz}$ for $\SI{10}{\milli\second}$ on the first red sideband only. The laser probe was set to a low power in order to resolve the COM and tilt modes (Rabi frequency on the carrier of $\sim\SI{14}{\kilo\hertz}$
). A clear asymmetry can be seen between the blue sidebands and the red sidebands, which are almost completely suppressed. A noticeable feature of the spectrum is that the tilt mode peak does not reach the height of the centre of mass peak and appears broader. This is most likely due to the instability of the rotation frequency of the crystal during the  acquisition of the spectrum. To account for this factor, and to take into account the effect of decoherence, the fitting function in the vicinity of the tilt mode and the centre of mass mode peaks was convoluted with a normalised Gaussian function, where the standard deviation was left as a free parameter.  The fitting model is the same as for the two ions in an axial string, but there is now no relative detuning between the two ions, and the Lamb-Dicke parameters of the two modes are the same.  
Since the fluorescence of each ion cannot be recorded individually for the radial crystal, we set the detection threshold to correspond to excitation of both ions simultaneously, because experimentally we can only reliably distinguish between both ions being dark and at least one ion fluorescing.  Note that the red sideband of both motions is very strongly suppressed because it is necessary for the mode to be in the state $n=2$ or higher for the red sideband to appear in the two-ion spectrum.  We find that the average phonon number for both modes is $\bar{n} < 0.15$ with a Gaussian broadening of the centre of mass mode and the tilt mode  of \SI{0.7}{\kilo\hertz} and \SI{1.0}{\kilo\hertz} respectively. From the position of the sidebands, the rotation frequency can be estimated using Equation \ref{freqEff} which yields {$\nu_r=\SI{106}{\kilo\hertz}$} (approximately 9\,kHz greater than the magnetron frequency). This rotation speed is consistent with measurements of the size of the image of the crystal.  
We observe a heating rate of 0.8(5)\,s$^{-1}$ for the COM  mode, consistent with our single ion rate. 

    
\begin{figure} 
\begin{center}
\includegraphics[width=0.77\columnwidth]{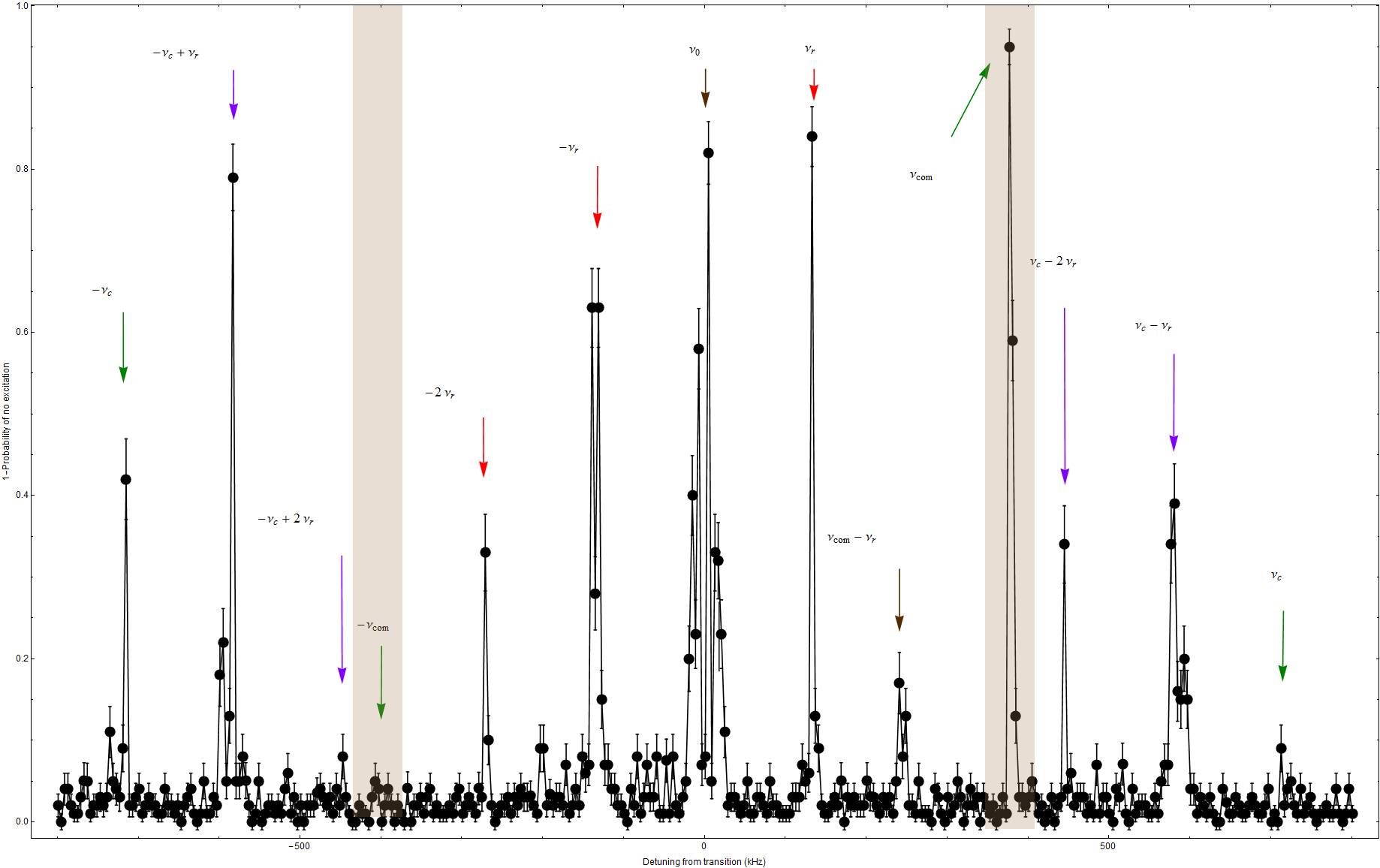}
\caption{Spectrum of a three-ion planar crystal after \SI{30}{\milli\second} of sideband cooling at a trapping frequency $\nu_z=\SI{379}{\kilo\hertz}$. A 4-V  axialisation drive was applied during cooling. The labels show the frequencies corresponding to the peaks. The ordinate represents the probability of excitation of at least one ion.  Note the asymmetry in the shaded peaks labelled $\nu_\text{com}$ and $-\nu_{\text{com}}$, which confirms that the system is cooled close to the ground state. The centre of mass and tilt modes are not resolved in this plot. }
\label{3ionPlanarSBC}
\end{center}
\end{figure}

A three-ion planar crystal was sideband cooled with a  sequence of three laser pulses on the first, second and first red sidebands respectively. Figure~\ref{3ionPlanarSBC} shows the spectrum obtained for the three-ion planar crystal after sideband cooling for a total of \SI{30}{\milli\second} (\SI{10}{\milli\second} on each pulse) at a trapping frequency $\nu_z=\SI{379}{\kilo\hertz}$.  It can be seen  that the amplitude of the red sideband of the COM mode is close to zero while the blue sideband almost reaches one. This large asymmetry suggests a high ground state occupation probability, although the complexity of the spectrum prevents us from fitting the data points and obtaining a more accurate figure. Due to the misalignment of the laser beam, many radial features can also be observed and identified on the spectrum.  {This includes sidebands on each of the main axial peaks due to the crystal rotation in the radial plane.  The  fact that the red radial sidebands are much more prominent than the blue ones implies a radial offset of the probe beam, as well as an angular misalignment}.
    
It is of interest to note that while the  ICCs in this paper have been cooled close to the quantum ground state in the axial direction, the tangential speed of the ions due to the bulk rotation of the ICC remains on the order of \SI
{10}{\meter\per\second} in the laboratory frame. This  demonstrates that the two bulk modes of this ion crystal, one of which is in the quantum mechanical ground state with low heating rate, have 
energies differing by four orders of magnitude, a situation we believe to be unique to this type of system.
    

\section{Conclusion}  

We have demonstrated the sideband cooling of small ICCs held in a Penning trap. Ground state confinement of all axial modes was achieved with high probability, for both linear and planar ICCs. When cooling multi-ion crystals from far beyond the Lamb-Dicke regime, sequential addressing of several sidebands including at least one intermodulation product is necessary to prevent population trapping.
Scaling up the number of ions, Penning traps are not well suited to confinement of much longer ion strings, as the very low axial trapping frequencies required make Lamb-Dicke confinement impossible. However, the techniques developed in this work could be readily extended to larger planar crystals or even three-dimensional ICCs, geometries which are difficult to obtain in a Paul trap. Cooling these structures to the Lamb-Dicke regime is a prerequisite for performing high fidelity entangling operations on ICCs and for producing strong and low-noise optical dipole forces in quantum simulators~\cite{Bohnet2016}. Such quantum simulations are typically performed on significantly larger ICCs of many hundreds of ions. Here the complexity of the mode structure may make EIT cooling~\cite{Lechner2016} a more practical means of achieving sub-Doppler temperatures, but resolved sideband techniques will remain useful for the direct thermometry and manipulation of individual motional modes. Given the richness of physics that has been demonstrated with simple one dimensional strings of ions, it is very interesting to consider the possibilities presented by cooling more complex structures to their motional ground state.
    
\section{Acknowledgements}  
This work was supported by the UK Engineering and Physical Sciences Research Council (Grant EP/D068509/1) and by the European Commission STREP PICC (FP7 2007-2013 Grant 249958). We gratefully acknowledge financial support towards networking activities from COST Action MP 1001 - Ion Traps for Tomorrow's Applications.  The research leading to these results has received funding from the People
Programme (Marie Curie Actions) of the European Union's Seventh
Framework Programme (FP7/2007-2013) under REA grant agreement no 31723. JFG acknowledges support from an EPSRC Doctoral Prize award. The optical system used for the images shown in Figure 1 and the spectra in Figures 4 and 5 was designed by Mr Jieyi Liu. 
    \bibliography{}

\end{document}